\documentclass[prb,aps,preprint]{revtex4}
\usepackage{graphicx}
\usepackage{epsfig}

\begin{document}

\title{Two-color pump-probe dynamics \\of transitions between doubly excited states of Helium}

\author{Spyros I. Themelis}
\email{stheme@iesl.forth.gr }
\affiliation{Institute of Electronic
Structure and Laser, Foundation for Research and Technology -
Hellas,\\P.O. Box 1527, GR-711 10 Heraklion, Greece}

\begin{abstract}
We discuss the dynamics of the two-photon resonant ionization of
helium involving two autoionizing states in the presence of
two-color laser fields. The first source is tuned around the
transition from the ground state to the 2s2p ${^1}P^o$
autoionizing state, and the second couples the state 2s2p
${^1}P^o$ to the $2p^2$~${^1}S$ autoionizing state. The laser
coupling between the doubly excited states is shown to lead to
modifications of the Beutler-Fano profile and the appearance of an
Autler-Townes doublet. This double resonance effect between
autoionizing states can be observed at moderate laser intensities
easily attainable by currently operated sources.
\end{abstract}

\maketitle

\section{INTRODUCTION}

The development of lasers of high-intensity and high-frequency
made possible the experimental investigations of many atomic
systems and revealed new phenomena. For example, the strong
electromagnetic coupling of two autoionizing states and their
coherent interaction, in some cases, could lead to partial
stabilization and population trapping in the ground
state~\cite{PL81}. The interaction of atomic autoionizing states
with an external electromagnetic field has been considered in a
number of papers~\cite{PL81,And82,Ba86,Lam89,TN99}. An
experimental and theoretical investigation of the effects of the
strong electromagnetic coupling of two autoionizing states on the
photoionization properties of Mg demonstrated the coherent
interaction between the autoionizing states~\cite{Kar95}.

Many experiments involving autoionizing states, have been performed
over the last years and the observed dynamics present very
interesting features. In this study, we consider the case of helium
atom in which the ground state is coupled to the 2s2p ${^1}P^o$
autoionizing state, through a laser with frequency $\omega_1$. A
second laser, with frequency $\omega_2$, couples the 2s2p ${^1}P^o$
state to the $2p^2~{^1}S$ autoionizing state. The latter has not
been observed in any photoexcitation experiment and the results of
the present study can be used for a new experimental investigation.
A similar system involving double autoionization resonance, has been
studied theoretically in detail, in connection with laser-induced
transitions between triply excited hollow states~\cite{LBM20}. Here
we present only the essential formulae of our approach, since the
the complete theoretical treatment has been published
elsewhere~\cite{PL81,LBM20}. The atomic parameters of the states
involved are calculated by an {\it ab-initio} approach. In Sec. 2 we
give a short description of our theoretical methodology and the
basic dynamic equations describing the problem. In Sec. 3 we solve
the equations and present and analyze our results. Finally, we
conclude in Sec. 4 with suggestions for prospective experimental
investigations.

\section{Theoretical approach}

We consider Helium atom subject to two linear polarized laser
fields with frequencies $\omega_1$ and $\omega_2$, respectively.
The relative phase between them is ignored and the total laser
field has the form:
\begin{equation}
E(t) = E_1 (t)\exp (i\omega _1 t) + E_2 (t)\exp (i\omega _2 t) +
c.c. .
\end{equation}
The frequencies $\omega_1$ and $\omega_2$ are chosen so as to be
tunable around the selected resonance transitions, namely
$1s^2~{^1}S \rightarrow 2s2p~{^1}P^o$ and $2s2p~{^1}P^o
\rightarrow 2p^2~{^1}S$. For the time-dependent field amplitudes
$E_i(t),~i=1,2$ we choose a convenient form for the pulse
envelope, namely a $\sin^2$, avoiding the long tails of a Gaussian
which make the numerics more difficult, without significantly
affecting the results. The explicit form is:
\begin{equation}
  E_i(t)=
  {\cal E}_i^{(0)}
  \sin^{2}(\frac{\pi t}{\tau _i} ),
  \quad
  \textrm{with}~
  0 \leq t \leq  {\tau _i},
\end{equation}
with ${\cal E}_i^{(0)}$ the the maximum field strength and
${\tau}_i/2$ is the Full Width at Half Maximum (FWHM). We assume
that ${\tau}_i$ is few picoseconds and the simultaneous action of
$E_i(t), i=1,2$, i.e. ${\tau}_1$=${\tau}_2$.

The next step is to solve the time-dependent Schr\"{o}dinger
equation. In the following sections we use the notation $ \left| g
\right\rangle$ for the ground state $1s^2~{^1}S$ of He, and
$\left| a \right\rangle$, $\left| {E_a } \right\rangle $ and
$\left| b \right\rangle$, $\left| {E_b } \right\rangle $ for the
discrete and continua parts belonging to the $2s2p~{^1}P^o$ and
$2p^2~{^1}S$ doubly excited states, respectively. Note that these
autoionizing states are single channel Feshbach resonances.  The
wave function for this standard model system shown in fig.1, can
be expressed as:
\begin{eqnarray}
\left| {\Psi (t)} \right\rangle &=& C_g(t)\left| g \right\rangle +
C_a(t)\left| a \right\rangle + C_b(t)\left| b \right\rangle \nonumber \\
&+& \int {dE_a C_{E_a } (t)\left| {E_a } \right\rangle } + \int
{dE_b C_{E_b } (t)\left| {E_b } \right\rangle } .
\end{eqnarray}
The Hamiltonian operator of the system is written as: $H=H_0 + V +
V_d$, with $ H_0\left| {\mu} \right\rangle = E_{\mu}\left| {\mu}
\right\rangle ,   ~\mu=a,b,g$, and $V$ being the configuration
interaction coupling the discrete parts of the doubly excited states
to the continua and $V_d = V_d(t)$ is the field-atom interaction.
Projection of the individual states in the expansion of $\left|
{\Psi (t)} \right\rangle$ leads to a set of coupled differential
equations containing amplitudes for the discrete parts as well as
for the continua. The introduction of (i) the slowly varying
amplitudes $c_{i} (t)$ which are defined by $c_{i} (t)$ = $C_{i}
(t)e^{i(E_i/\hbar+\Delta \omega)t}$  and $\Delta \omega$ is the sum
of the frequencies of the absorbed photons, (ii) the application of
the rotating wave approximation(RWA), and (iii) the adiabatic
elimination of the continua lead to the following set of equations
for the discrete-state amplitudes:
\begin{equation}
{i \hbar} \frac{{\partial }}{{\partial t}} {\bf c}(t) = {\bf H}(t)
{\bf c}(t),
\end{equation}
where
\begin{equation}
{\bf H}(t) = \left[
\begin{array}{ccc}  {S_g  - \frac{i}{2}\gamma _g }  & \tilde \Omega _{ga} &
{S_{gb}  - \frac{i}{2}\gamma _{gb} }   \\
\tilde \Omega_{ag} &  -\delta _1  - \frac{i}{2}(\Gamma _a  +
\gamma _a )  & \tilde \Omega _{ab} \\
{S_{bg}  - \frac{i}{2}\gamma _{bg} }  & \tilde \Omega _{ba} &
-{\delta _1  - \delta _2  - \frac{i}{2}(\Gamma _b  + \gamma _b )}
\end{array} \right],
\label{eq:H}
\end{equation}
and ${\bf c}(t) = \left[ c_{g}(t), c_{a}(t), c_{b}(t) \right]^T$.
In equation (5) $ \delta _1 = \omega _1  - (E_a^{(0)}  + S_a  -
E{}_g - S_g) $, $ \delta _2 = \omega _2  - (E_b^{(0)}  + S_b  -
E_a^{(0)}  - S_a )$ are the detunings, $E_a^{(0)}$, $E_b^{(0)}$
and $\Gamma _a$, $\Gamma _b$ the resonance energy and width of the
doubly excited states, $S_g$, $S_a$, $S_b$, $S_{gb}$ and $\gamma
_g$, $\gamma _a$, $\gamma _b$, $\gamma_{gb}$ the laser induced
shifts and widths. The generalized complex Rabi frequencies
$\tilde \Omega _{ga} $ and $\tilde \Omega _{ab}$ are defined as:
\begin{equation}
\tilde \Omega _{ga} = \Omega _{ga} \left( {1 - \frac{i}{{q_a }}}
\right) = \frac{1}{2}E_1 (t)D^{(z)}_{ga} \left( {1 - \frac{i}{{q_a
}}} \right)
\end{equation}
and
\begin{equation}
\tilde \Omega _{ab} = \Omega _{ab} \left( {1 - \frac{i}{{q_{ab}
}}} \right) = \frac{1}{2}E_2 (t)D^{(z)}_{ab}\left( {1 -
\frac{i}{{q_{ab} }}} \right)
\end{equation}
with $D^{(z)}$ being the dipole along the polarization direction
$z$, and $q_a$ and $q_{ab}$ the Fano line shape parameter for the
transition $\left| g \right\rangle \rightarrow \left| a
\right\rangle$  and its generalization for the transition  $\left|
a \right\rangle \rightarrow \left| b \right\rangle$. Obviously the
Hamiltonian of equation (\ref{eq:H}) is \emph{Non-Hermitian}. The
coefficients $c_{i}(t)$ are slowly varying in the sense that the
transformation $c_{i} (t)$ = $C_{i} (t)e^{i(E_i/\hbar+\Delta
\omega)t}$ has removed their rapid variation.

\begin{figure}
\includegraphics[width=12cm]{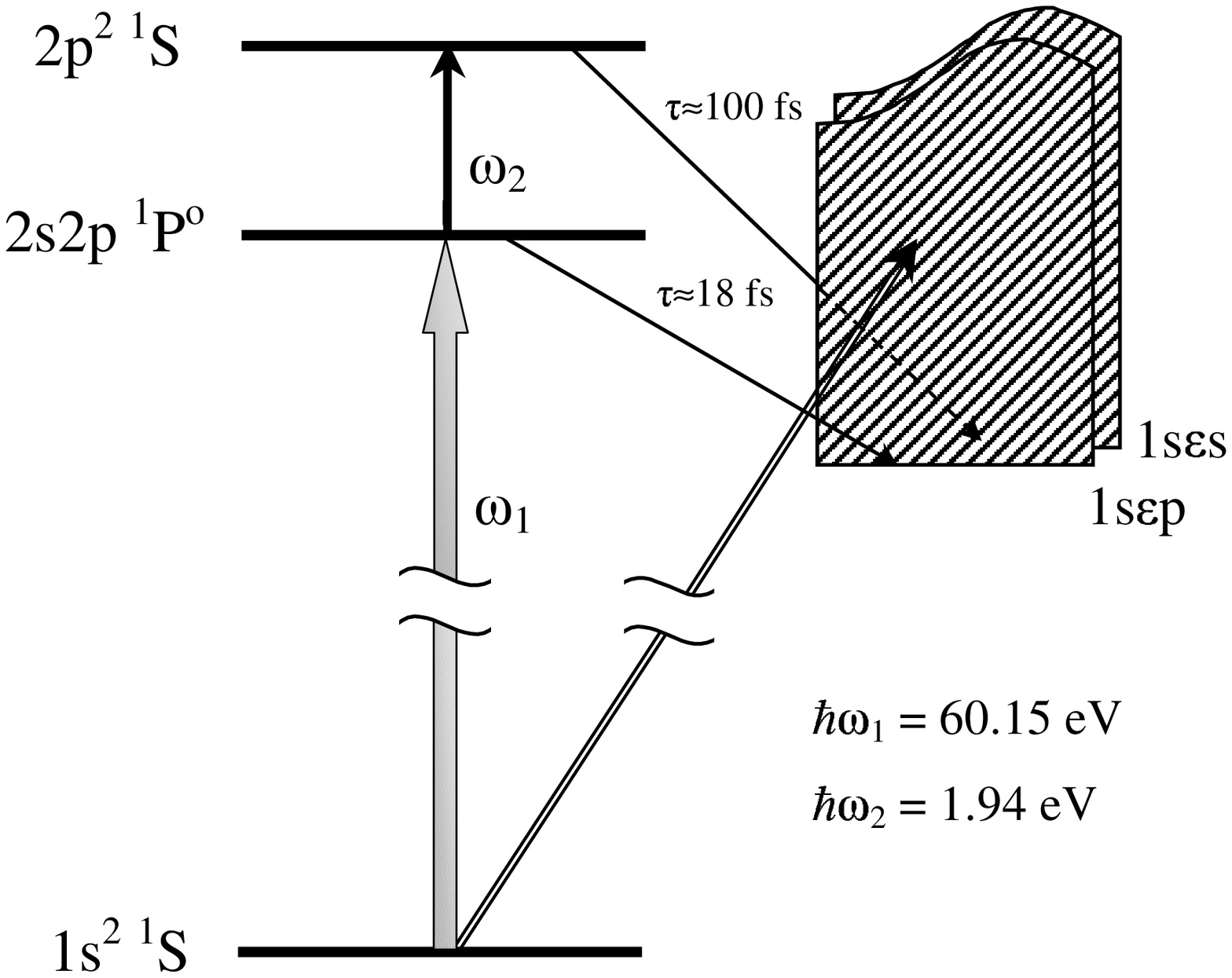}
\caption{Schematic representation of the coupling between the
ground, the two autoionizing states and the continua entering the
problem.}
\end{figure}
For the derivation of the above equations the laser induced
continuum-continuum couplings have been neglected. It has been
shown that the line shapes are not affected by continuum-continuum
transitions and the same is true and for the total photoionization
rate~\cite{Cav95}. Also, preliminary calculations showed that we
can ignore the laser-induced couplings between the discrete parts
$\left| a \right\rangle$ and $\left| b \right\rangle$ of the
resonance states and the non-resonant part of the continua $\left|
E_b \right\rangle$ and $\left| E_a \right\rangle$ respectively,
and the second order effect of the laser induced coupling of the
ground state $\left| g \right\rangle$ to $\left| b \right\rangle$
via the non-resonant part of the continua $\left| E_a
\right\rangle$. By solving the above equations the ionization
yield into each channel and the total ionization probability can
be calculated. The total ionization probability is:
\begin{equation}
P{\rm{ (}}t{\rm{) }} = {\rm{ 1 }} - \left| {{\rm{c}}_{\rm{g}} (t)}
\right|^2  - \left| {{\rm{c}}_{\rm{a}} (t)} \right|^2  - \left|
{{\rm{c}}_{\rm{b}} (t)} \right|^2
\end{equation}
\begin{table}
\begin{tabular}{ccccccccc}
\hline \hline ~~~$q_a$~~~ & ~~~$q_{ab}$~~~&
~~~~$E_a^{(0)}$~~~~&~~~~ $\Gamma _a$~~~~&$~~~~E_b^{(0)}$~~~~ &
~~~~$\Gamma _b$~~~~
&~~~~$\Omega _{ga}$~~~~ & ~~~~$\Omega _{ab}$~~~~ & ~~~~$\gamma _g$~~~~  \\
\hline
 -2.79&-714 &-0.6928&1.37 $\times 10{^{-3}}$ &-0.6214 &
2.15$\times 10{^{-4}}$
& $ 0.038 \frac{E_{1}(t)}{2} $ & $ 2.14 \frac{E_{2}(t)}{2} $  & $ 0.47 \frac{I_{1}(t)}{4} $  \\
\hline
\hline
\end{tabular}

\caption{Coupling of the states involved in the 3$\times$3 model
examined here. All values are given in atomic units. $E_a^{(0)}$,
$E_b^{(0)}$ and $\Gamma _a$, $\Gamma _b$  are the energies and the
widths of the 2s2p${^1}P^o$ and $2p^2~{^1}S$ states,
respectively.}
\end{table}
{\emph{Atomic structure parameters}}. We have calculated all of the
parameters pertaining to the atomic levels coupled by the process
described above, through an {\it ab initio} approach. In order to
take into account the electron correlation, very accurately, a large
number of configurations were selected. The MCHF
method~\cite{CFF2000} has been used to perform the present
calculations. The MCHF wave-function expansion for the ground state
of He was over a set of 15 configuration states coupled to form a
$^{1}S$ term. The radial wave functions for the different orbitals
were obtained by the MCHF procedure, varying all the orbitals
simultaneously. Minimization of the total energy yielded an energy
of -2.9033 a.u. to be compared to the accurate value of -2.903724
a.u. from extensive variational calculations. The autoionizing
states 2s2p ${^1}P^o$ and $2p^2$~${^1}S$ of He are calculated using
a partition of the function space within the framework of the
Feshbach formalism~\cite{Fes62}. Using appropriate $\it Q$ and $ \it
P$ projection operators, we can represent these resonances as
quasibound states embedded in a continuum. The localized
wavefunctions  and  of these states are also calculated by a MCHF
approach. For 2s2p ${^1}P^o$ we used 27 configuration states and the
energy obtained for this state is: -0.69256 a.u. For the
$2p^2$~${^1}S$ state we used 30 two-electron configuration states
and the energy obtained for this state is: -0.62218 a.u. The
correlation effects are important for an accurate description of
these states. This can be seen from their configuration expansion,
which for the 2s2p ${^1}P^o$ is of the form:
\[
\psi(2s2p~ {^1}P^o) =  0.953 (2s2p) - 0.291 (2p3d) - 0.076 (3s3p)
+ ...
\]
while for the $2p^2$~${^1}S$ state we have:
\[
\psi(2p^{2}~{^1}S) = 0.787 (2p^2) - 0.554 (2s^2) + 0.173 (3s^2) -
0.139 (3p^2) + 0.179 (3d^2) + ...
\]
The autoionization widths of these states are calculated by the
well known complex-coordinate method~\cite{Byl98}. We choose the
open channel component of the resonant wavefunctions to be:
\begin{equation}
u(1s\varepsilon \ell~ ^{1}L^{\pi}) = \hat{ \cal A} \left( \phi_
{1s}(r_1) \sum_{i} c_i \chi_i ( \rho_2 ^{*}) Y_{\ell m}(\Omega_2)
\right)
\end{equation}
with $\chi_i ( \rho ^{*}) = (\rho ^{*})^{k_i} e^{-a_i \rho ^{*}}$.
The radial function $\phi_ {1s}(r)$ was kept fixed to the
hydrogen-like orbital of He$^+$ whereas, in the variant of the
complex-coordinate approximation followed here, for $\chi_i ( \rho
^{*})$ the radial coordinate takes the form $\rho_i ^{*}=r_i
e^{-i\theta}$. The non-linear parameters $a_i$ and the expansion
coefficients $c_i$ are subject to a variational optimization for
the calculation of the complex energy eigenvalues pertaining to
the autoionizing resonant states.

Our results for the energy position, including the energy shift,
and the width, of the autoionizing states presented here, can be
compared to other more elaborate calculations~\cite{Lind94}. The
wave-functions described above were also used for the calculation
of the dipole moments for the various transitions involved. All
the parameters that enter in the calculation are shown in table 1.
We note here that the use of the complex-coordinate method
provides a powerful tool for the calculation of the complex Rabi
frequency $\tilde \Omega _{ab}$ for a transition between
autoionizing states.
\section{Results and discussion}
\begin{figure}
\includegraphics[width=10cm]{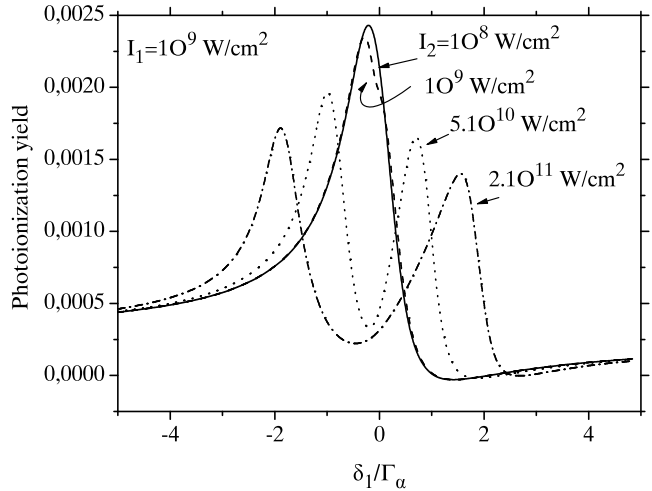}
\caption{He$^+$ photoion yield as a function of the detuning
${\delta}_1$, associated with the transition between the ground
state and the 2s2p ${^1}P^o$ doubly excited state. The laser
coupling the 2s2p ${^1}P^o$ and $2p^2$~${^1}S$ states is on
resonance (${\delta}_2$=0). We clearly see that the laser coupling
induces an Autler-Townes doublet.}
\end{figure}

The solution of the system of differential equations given by Eq.
(4) can provide us the information about the temporal evolution of
the system under consideration. We have chosen to study the
response of the system under the simultaneous action of the
electric field $E_1$ with frequency $\omega_1$ and a pulse
duration ${\tau _1}$=5 ps and of the electric field $E_2$ with
frequency $\omega_2$ and of the same duration.

Figures 2 and 3 show the photoionization yield of He$^+$ as a
function of the detunings of the laser sources for a series of
intensities and detunings. As it can be seen from Fig. 2 the line
shape changes significantly with the intensity of the laser
coupling the 2s2p ${^1}P^o$ and $2p^2$~${^1}S$ doubly excited
states. At low intensities we have a line shape for the lowest
${^1}P^o$ Feshbach resonance of He which is a typical Beutler-Fano
profile~\cite{Fan61}. As the intensity is increased a doublet
appears due to the ac Stark splitting as a result of the laser
induced oscillation between 2s2p ${^1}P^o$ and $2p^2$~${^1}S$.
This structure is known as an Autler-Townes doublet and the
separation between the two peaks carries information about the
dipole matrix element coupling the doubly excited states. In Fig.
3 the frequency $\omega_1$ is on resonance (${\delta}_1$=0) while
the laser frequency $\omega_2$ is varied (${\delta}_2 \ne 0$). In
this case the coupling of the doubly excited states reveals a
window resonance on the photoionization cross section.
\begin{figure}
\includegraphics[width=10cm]{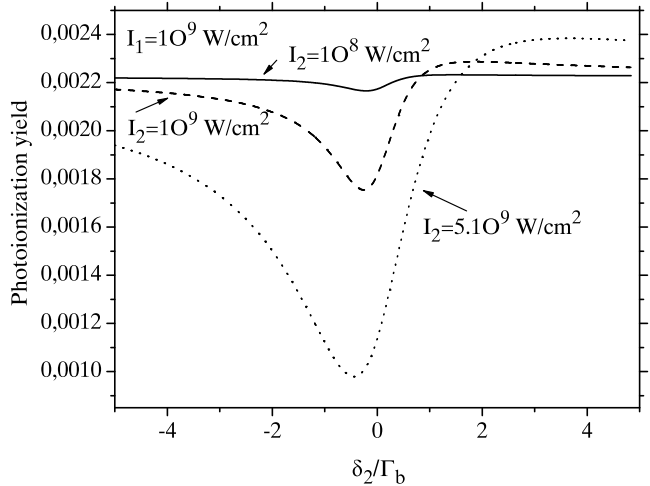}
\caption{He$^+$ photoion yield as a function of the detuning
${\delta}_2$, associated with the 2s2p ${^1}P^o$ $\rightarrow$
$2p^2$~${^1}S$ transition. The laser coupling the ground state
with the 2s2p ${^1}P^o$ doubly excited state is on resonance
(${\delta}_1$=0). We clearly see that the appearance of a window
resonance as the laser intensity increases.}
\end{figure}

The number of the electrons emitted in the energy region of each
autoionizing state is proportional to the ionization signal given
by the formula:
\begin{equation}
S_i = \int\limits_{ - \infty }^{ + \infty } {\left| {c_i (t)}
\right|^2 \Gamma _i dt}
\end{equation}
with $i=a, b$. The magnitude of $S_i$ depends on the time
evolution of the coefficients $c_i$ and, in a more crucial way, on
the values of the autoionizing widths of the resonant states. We
have calculated $S_a$ and $S_b$ for various values of $I_2$
ranging from $10^8 W/cm^2$ to $2 \times 10^{11} W/cm^2$ and for
(${\delta}_1 \ne 0$, ${\delta}_2=0$) and our results are shown in
fig. 4a and 4b.
\section{CONCLUSIONS}
To our knowledge the autoionizing state $2p^2$~${^1}S$ of He has not
been observed in any multiphoton process. There has not been any
detection of this state by two-photon processes and the energy
position of this resonance is measured only in scattering
experiments. In this study we showed that, the coupling of this
state via a laser field of moderate intensity to another
autoionizing state, has significant and detectable results in the
photoionization cross section of He. The prospective pump-probe two
color experiments utilizing sources as, for example, the Deutsches
Elektronen-Synchrotron Free Electron Laser (DESY FEL) provide an
apparatus for the detection of the resonant coupling of autoionizing
states as these described here~\cite{Mey01}. The laser intensities
used in our study are attainable and proposed experiments with
Free-Electron Laser (FEL) sources are readily
available~\cite{Andr00,Fel03}.

\begin{figure}
{\bf (a)} \includegraphics[width=7.cm]{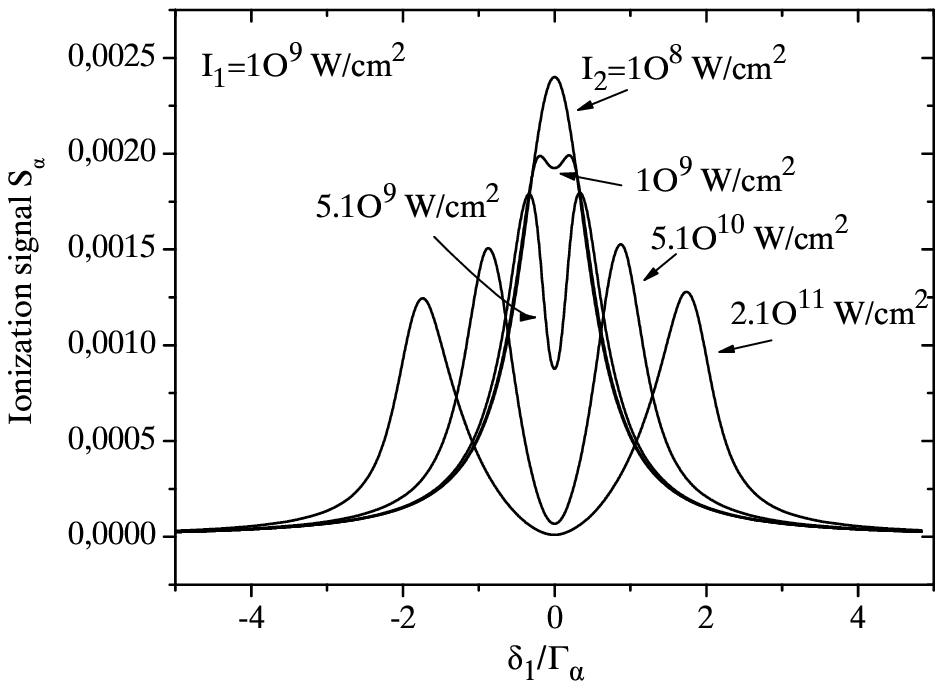} \hskip0.1cm {\bf
(b)}\includegraphics[width=7.cm]{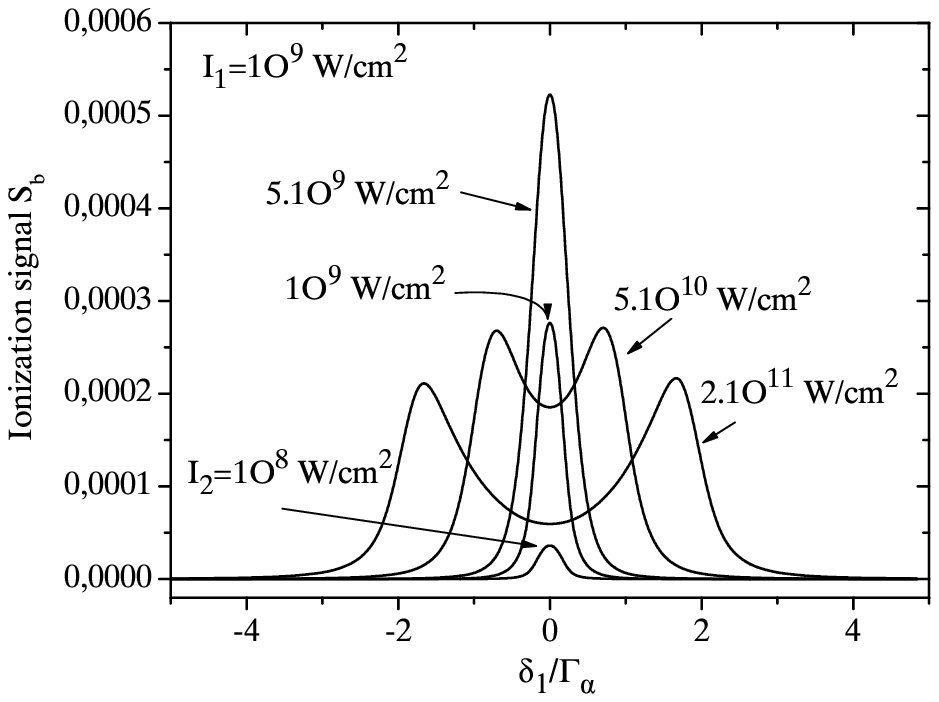} \caption {The
ionization signal in the energy range of the autoionizing states
(a) 2s2p ${^1}P^o$ and (b) $2p^2~{^1}S$, for various values of the
laser intensity $I_2$. The laser coupling the 2s2p ${^1}P^o$ and
$2p^2$~${^1}S$ states is on resonance (${\delta}_2$=0).}
\end{figure}

\begin{acknowledgments}
The author is indebted to Prof. P. Lambropoulos for fruitful and
inspiring discussions which initiated the project.

\end{acknowledgments}


\begin{thebibliography}{90}

\bibitem{PL81} Lambropoulos P and Zoller P 1981 {\it Phys. Rev.} A {\bf 24}, 379

\bibitem{And82} Andryushin A I, Fedorov M V, and Kazakov A E 1982
{\it J. Phys. B: At. Mol. Phys.} {\bf15} 2851

\bibitem{Ba86} Bachau H, Lambropoulos P, and Shakeshaft R 1986 {\it Phys. Rev.} A {\bf34} 4785

\bibitem{Lam89} Lami A, Rahman N K, and Spizzo P 1989 {\it Phys.
Rev.} A {\bf40} 2835

\bibitem{TN99} Nakajima T 1999 {\it Phys. Rev.} A {\bf60} 4805

\bibitem{Kar95} Karapanagioti N E, Faucher O, Shao Y L, Charalambidis D, Bachau H,
and Cormier E 1995 {\it Phys. Rev. Lett.} {\bf 74} 2431

\bibitem{LBM20} Madsen L B, Schlagheck P, and Lambropoulos P 2000 {\it Phys. Rev.
Lett.} {\bf85} 42

\bibitem{Cav95} Cavalieri S, Eramo R, Buffa R, and Matera M 1995 {\it Phys. Rev.} A {\bf51}
 2974

\bibitem{CFF2000} Froese Fischer C 2000 {\it Comput. Phys.
Commun.} {\bf128} 635

\bibitem{Fes62} Feshbach H, 1962 {\it Ann. Phys.} (N.Y.) {\bf 19} 287

\bibitem{Byl98} Bylicki M 1998 {\it Adv. Quant. Chem.} {\bf32}
207

\bibitem{Lind94} Lindroth E 1994 {\it Phys. Rev.} A {\bf49} 4473

\bibitem{Fan61} Fano U 1961 {\it Phys. Rev.} {\bf124} 1866

\bibitem{Mey01} Meyer M, Gisselbrecht M, Marquette A, Delisle C, Larzill\`{e}re M, Petrov
I, and Sukhorukov V 2001 {\it Nucl. Inst. and Meth.} A
{\bf467-468} 1447

\bibitem{Andr00} Andruszkow J, {\it et al.}, 2000 {\it Phys. Rev. Lett.} {\bf 85}
3825

\bibitem{Fel03} Feldhaus J, M\"{o}ller T, Saldin E E, Schneidmiller E A, and Yurkov M
V 2003 {\it Nucl. Inst. and Meth.} A {\bf507} 435

\end{thebibliography}
\end{document}